\documentstyle[12pt,ccgrra]{article}
\begin{document}

\include{psfig}

\newcommand{\R}{{\mathbb R}}
\newcommand{\C}{{\mathbb C}}
\newcommand{\Z}{{\mathbb Z}}   
\newcommand{\be}{\begin{equation}}
\newcommand{\ee}{\end{equation}}
\newcommand{\bea}{\begin{eqnarray}}
\newcommand{\eea}{\end{eqnarray}}
\newcommand{\tb}{\tilde{\beta}}
\newcommand{\ta}{\tilde{a}}
\newcommand{\tal}{\tilde{\alpha}}
\newcommand{\ttau}{\tilde{\tau}}
\newcommand{\tp}{\tilde{p}}
\newcommand{\cQ}{{\mathcal{Q}}}
\newcommand{\cF}{{\mathcal{F}}} 
\newcommand{\cG}{{\mathcal{G}}}
\newcommand{\cH}{{\mathcal{H}}}
\newcommand{\cX}{{\mathcal{X}}}
\newcommand{\cY}{{\mathcal{Y}}}
\newcommand{\cR}{{\mathcal{R}}}
\newcommand{\cA}{{\mathcal{A}}}  
\newcommand{\La}{\frac{\Lambda}{3}}
\newcommand{\nn}{\nonumber}

\title{COSMOLOGICAL PAIR CREATION OF CHARGED AND ROTATING BLACK HOLES}

\author{I. S. N. Booth\dag \footnote{e-mail: ivan@avatar.uwaterloo.ca},
R. B. Mann\dag \footnote{e-mail: mann@avatar.uwaterloo.ca}}

\affil{\dag Dept. of Physics,
University of Waterloo, Waterloo, ON   \\N2L 3G1, CANADA}

\beginabstract
In this paper, we investigate the creation of charged and rotating
black hole pairs in a background with a positive cosmological constant.
Instantons to describe this situation are constructed from the
Kerr-Newmann deSitter solution to the Einstein-Maxwell equations and the
actions of these instantons are calculated in order to estimate pair
creation rates and the entropy of the created spacetimes.
 \endabstract

\section{Introduction}
In recent years there has been a considerable amount of interest in black
hole pair production. Inspired by the more mundane and well understood
particle pair production of quantum field theory (for example $2 \gamma
\rightarrow e^+ + e^-$), studies have been made to investigate the
probability that a spacetime with a source of excess energy will quantum
tunnel into a spacetime containing a pair of black holes. The earliest
investigations studied pair creation due to background electromagnetic
fields \cite{dgk} but
since then people have studied pair creation powered by domain walls
\cite{ccg}, cosmic strings \cite{string}, and the cosmological constant
\cite{rob}. To
date, the studies have focussed on the creation of static black holes. In
this paper we seek to extend the work on cosmological constant
pair creation to include the creation of rotating black holes.

As in the extant studies, we shall conduct this investigation within the
framework of the path integral formulation of quantum gravity, of which we
now present a lightning review. The standard problem of quantum
mechanics is to calculate the
probability that a system passes from an initial state $X_1$ to a final
state $X_2$. If the system in question is a four dimensional spacetime,
then these states are each characterized by a three
manifold $\Sigma$, with Riemannian metric $h_{ij}$, and a symmetric
tensor field $K_{i j}$ (the extrinsic curvature of the surface) which
describes how the three manifold is embedded
in the wider space time (or on a more physical level, roughly describes
how the three manifold is evolving at that ``instant'' in time).
Matter fields may be present on the spacetime in which
case their values on the three manifold must also be specified. Then, in
the path integral approach, we consider \underline{all} four manifolds,
$M$,  and
metrics and fields, $g$ and $A$, on those manifolds that smoothly
interpolate between the
initial and final
conditions (not just those that are solutions to the equations of motion).
The probability
amplitude that a space time evolves from the initial condition $X_1$ to a
final condition $X_2$ may be written as a functional integral over all
such manifolds,
\be
\Psi_{12} = \int d[M]d[g]d[A] e^{-i I(M, g, A)},
\ee
where $I$ is an appropriate action for the situation being studied. In
this context, appropriate means an action that may be varied to give the
correct set of field equations for classical motion plus an acceptable set
of fixed quantities
on the boundaries of the regions $M$ (for a more complete discussion, see
section \ref{action}).
In analogy with flat-space calculations, it is then argued
\cite{origact} that
to lowest order, this amplitude may be approximated by,
\be
\label{pairc}   
\Psi_{12} \approx e^{-I_c}
\ee
where $I_c$ is the real action of a (not necessarily real) Riemannian
solution to the Einstein-Maxwell equations
that smoothly interpolates between the given initial and final conditions.
Such a solution is referred to as an instanton.
     
With this review of the path integral formalism in mind, we now turn to
a preview of how it will be applied to the work at hand. First off, we
need a metric that describes a deSitter spacetime containing a pair of
charged and rotating black holes. These requirements are fulfilled by
the Kerr Newmann deSitter metric, which we present in section
\ref{KNdSsect}. 
Secondly, an instanton solution is needed that interpolates between this
solution and some initial condition. In section
\ref{construct} we construct such a solution from the KNdS solution,
and along the
way choose our initial condition to be the no boundary condition. With 
the instanton in hand, we must then calculate its action in order to
estimate the probability of pair creation and as a bonus the entropy of
the spacetimes. This will be the subject of section \ref{action}.
 
\section{The Kerr-Newmann deSitter Metric}
\label{KNdSsect}
The Kerr-Newmann deSitter solution to the Einstein-Maxwell equations
has the following form \cite{MM}.
\bea
\label{KNdS}
ds^2  =  -\frac{\cQ}{\cG \chi^4} \left(  dt - \alpha \sin^2\theta d\phi
\right)^2 + \frac{\cG}{\cQ} dr^2
      + \frac{\cG}{\cH} d\theta^2 + \frac{\cH \sin^2 \theta}{\cG \chi^4}
\left( \alpha dt - \left[ r^2 + \alpha^2 \right] d\phi \right)^2,
\eea
where, $\cG = r^2 + \alpha^2 \cos^2 \theta$, $\cH = 1 +
\frac{\Lambda}{3} \alpha^2 \cos^2 \theta$, $\chi^2 = 1 + \frac{\Lambda}{3}
\alpha^2$,
\be 
\label{Qpoly}
\cQ = -\frac{\Lambda}{3} r^4 + \left( 1 - \frac{\Lambda}{3} \alpha^2
\right)
r^2 - 2Mr + \left( \alpha^2 + E_0^2 + G_0^2 \right), 
\ee
and the electromagnetic field is,
\be
\label{EMfield}
F = d \left\{ \frac{E_0 r}{\cG \chi^2} \left[ dt
 - \alpha \sin^2 \theta d\phi \right] + \frac{G_0
\cos \theta }{\cG \chi^2} \left[ dt - \left(
r^2 + \alpha^2 \right) d\phi \right] \right\}.
\ee
In this metric, $\Lambda$ is the cosmological constant (which for deSitter
solutions will be assumed to be positive), $\alpha$ is the rotation
parameter, $M$ parameterizes the mass, and $E_0$ and
$G_0$ are respectively the effective electric and magnetic charge of the
solution. The roots of $\cQ$ determine the locations of the black hole and
cosmological horizons. The absence of the cubic term in $\cQ$ means that
$\cQ$ will have at most three real and positive roots. We shall assume
that three such roots exist. In
ascending order they are the inner, outer, and cosmological horizons.  

\begin{figure}
\centerline{\psfig{figure=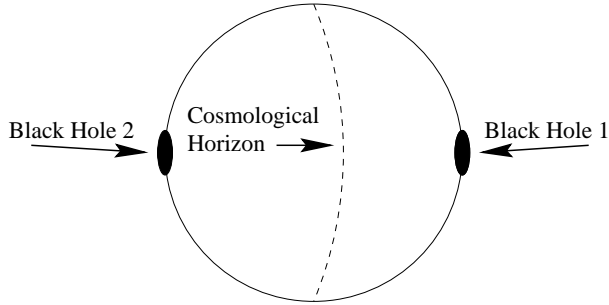,height=4cm,angle=270}}
\caption{A schematic of two KNdS metrics 
attached at their cosmological
horizons.}  
\label{fig1}
\end{figure}

In the context of pair creation studies, it is important to note that this
metric, which apparently describes a single black hole in deSitter space,  
may be analytically continued through the cosmological horizon to
describe a pair of black holes
sitting at ``opposite ends'' of the  universe, and separated by the
cosmological horizon. Essentially two identical such solutions are
joined together by identifying their cosmological horizons
(figure \ref{fig1}). For computational reasons, it will later be necessary
to  
formally identify their outer horizons (figure \ref{fig2}) as well. These
analytic extensions are discussed further in \cite{brill}.

\section{Constructing the Instantons}
\label{construct}

\begin{figure}[t]
\centerline{\psfig{figure=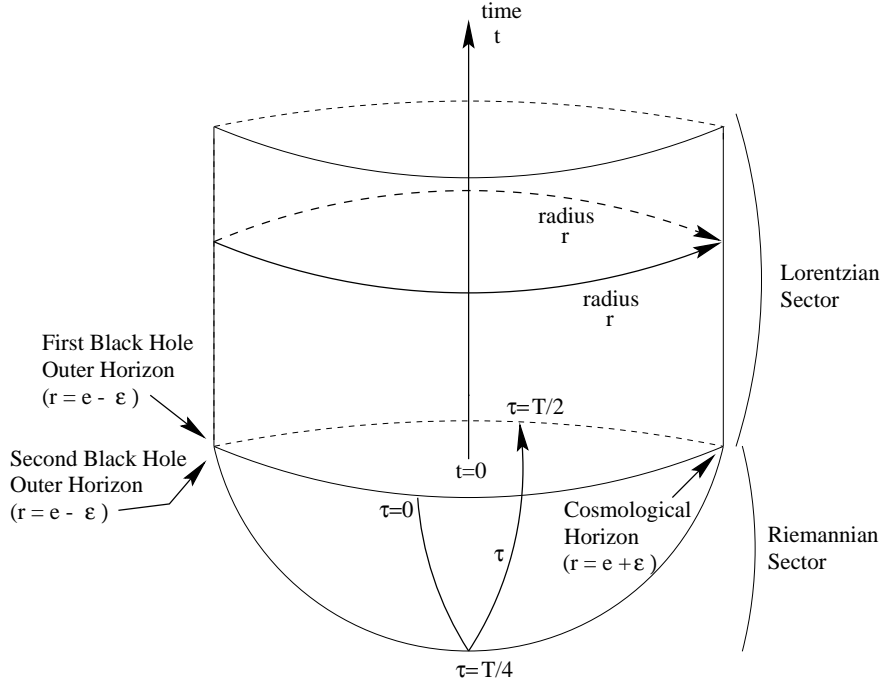,height=9cm,angle=270}}
\caption{The KNdS instanton cut in half, and attached to the two hole
KNdS solution. The view is of the $ r - \tau /t$ sector.}
\label{fig2}
\end{figure}

An instanton with complex metric may be constructed from the
single hole version of the KNdS
solution by restricting the radial coordinate
to lie between the outer black hole and cosmological horizons, sending $t
\rightarrow i \tau$ and periodically identifying the new imaginary time
with some period $T$. One half of this instanton may then be
wedded to the two hole KNdS solution, in order to describe pair
creation within the context of the no-boundary condition (see
figure \ref{fig2}). 

Things are not quite that simple however. In the previous paragraph, we
ignored the possibility that conical singularities can, and usually
will, show up in the metric if the period $T$ of $\tau$ is chosen in an
arbitrary
way. In order to avoid such singularities, the period must be chosen
with some care, and even then fairly strong additional restrictions must
be placed on the metric.

Briefly, conical singularities are defined in the context of two-manifolds
with $SO(2)$ symmetry groups that have a fixed point under the group
action. In
that context, the two-manifold may be foliated into a
family of concentric ``circles'' generated by the action of the group.
This family may be definitely ordered, and so a coordinate system may be
set up, with a coordinate radius $\zeta$ labelling which ``circle'' a
point is on, and an angular coordinate giving the position on the
``circle''. Then, bringing the metric into play, a physical radius
$R(\zeta)$ and circumference $C(\zeta)$ for each circle may be calculated. 
If $\lim_{\zeta \rightarrow 0} R = 0$ (ie the point is not a point 
at infinite proper distance from all other points), and $\lim_{\zeta
\rightarrow 0} \frac{C}{R} \neq 2 \pi$, then a conical singularity is said
to exist at the fixed point.

If we look at $r-\tau$ slices of the instanton metric, then there are
potential conical singularities at the cosmological and outer black hole
horizons (they are fixed points under $\tau$ translations). If the inner
black hole horizon coincides with the outer, then these lie at a double
root of $\cQ$ and it is not hard to see that the horizon is at an infinite
distance from all other points of the manifold, and so is not a
candidate for a conical
singularity. If we consider a single root however,
calculation shows 
that a conical singularity exists unless,
\be
\label{per}
T = 4\pi \chi^2 (r_i^2 + \alpha^2) / Q'(r_i),
\ee
where $r_i$ is the radial coordinate of the root.

Thus, we can see that there are two ways to eliminate the conical
singularities. Either we need a double root at the black hole horizon, in
which case we can adjust the period of $\tau$ to match that required by
the cosmological horizon,
or
we need to pick the physical parameters in $\cQ$ such that both the black
hole horizon and cosmological horizons are single roots, and have a common
required period. The first case will pair create extreme black holes.
We label such instantons as cold instantons in analogy
with
the equivalent Reissner-N\"{o}rdstom deSitter instanton \cite{rob}. There
are two ways to satisfy the second condition. The first provides distinct
horizons, while the second has the cosmological and black hole horizons
coincident. These are labelled as luke-warm and Nariai instantons, again
in analogy with the RNdS instantons \cite{rob}.  

\begin{figure}
\centerline{\psfig{figure=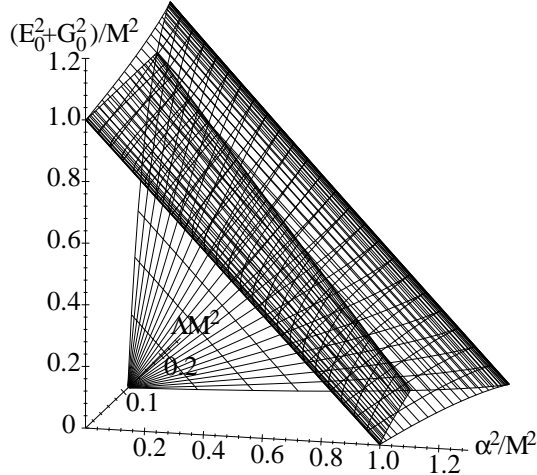,height=7cm}}
\caption{Allowed range of the KNdS instanton parameters. The closest
sheet represents the cold instantons, the middle the lukewarm, and
the farthest the Nariai.} \label{fig3} \end{figure}

\begin{figure}
\centerline{\psfig{figure=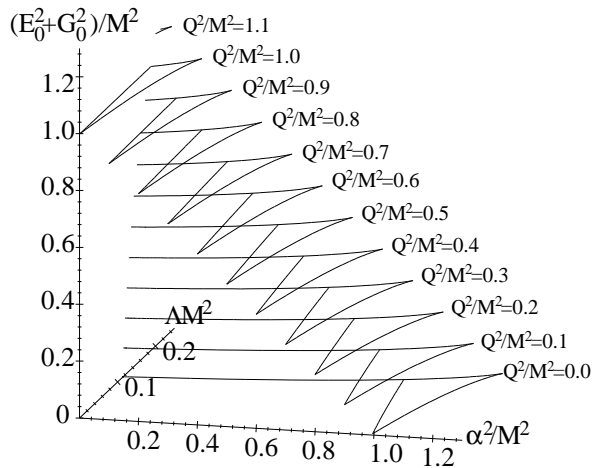,height=7cm}}
\caption{Contour plot of the allowed range of KNdS instanton parameters.}
\label{fig4}
\end{figure}

Figures \ref{fig3} and \ref{fig4} show the allowed range of the physical
parameters for each type of instanton. Due to space constraints, the
calculations leading up to these diagrams are omitted here. They will be
included in \cite{tbp}.

The proper physical interpretation of the Nariai instantons is not so
clear as for
the other two types of instantons. In fact, 
at first inspection they appear to be degenerate. They may be
rescued from degeneracy and to physical relevancy by
considering a combination of a mathematical limiting process and a quantum
correction as was originally done in
\cite{gp} and discussed in more detail in \cite{bousso}. Due to spatial
constraints, we will omit further consideration of the Nariai metric in
this summary.  

Finally, before moving on to a calculation of the actions, we should note
that by eliminating conical singularities from the instantons we have
equivalently ensured that the cosmological and black hole horizons have
the same surface gravity and temperature in the Lorentzian counterpart
solutions. That is, we have ensured that the created space time will be in
thermal equilibrium. Equivalently, there is no net particle flux between
the two horizons. In the
absence of charge and rotation, the system would then be in true
thermodynamic equilibrium and would manifest no time dependence. With
charge and/or rotation however, the black hole will evolve in time by
discharging and/or spinning down.

We have neglected both discharge and spin-down effects for the classes
of black holes we consider.  Discharge effects are small if the
temperature of a black hole is small relative to the mass of the
lightest charged particle. Spin-down effects resulting from the
quantum evaporation of spinning particles from the hole will also tend to be
small for black holes in thermal equilibrium with the cosmological
horizon. However for massless bosons of angular momentum $m$ with 
frequencies $\omega$
in the range $m\Omega_C < \omega < m\Omega_H$ where
$\Omega_{C,H} = \frac{\alpha}{r^2_{C,H}+\alpha^2}$ is the angular
velocity of the
cosmological/outer black hole horizon, these effects will be amplified by
superradiance. The relative rate of angular
momentum loss due to scalar fields for near extremally rotating,
uncharged, Kerr deSitter holes (not in thermal equilibrium with the
cosmological horizon) as amplified by superradiance, is $<\approx 2\times
10^{-4}$ \cite{maeda}, and is expected to be larger for bosons of
higher spin, increasing by as much as 2 orders of magnitude for graviton
emission. The effect is strongest for rapidly rotating black holes, but
decreases rapidly as the angular momentum parameter of the black hole
decreases (at least in the asymptopically flat case \cite{page}).
Now, uncharged lukewarm and cold holes are necessarily
rapidly rotating (see figures \ref{fig3} and \ref{fig4}),
but with the inclusion of charge
they need not be (figures \ref{fig3} and \ref{fig4} again), and in fact
they may rotate at arbitrarily slow rates.
Thus, it appears that while neither
the cold nor lukewarm holes are in complete thermodynamic equilibrium, at
least some subset of them will be close enough to
equilibrium that we may treat them as being in that state within the
approximations inherent in the path integral techniques.
For the Nariai spacetime, $r_C=r_H$ and so the spin down effect is absent.
A more complete discussion of these issues will appear in a forthcoming 
paper \cite{tbp}.

\section{Action and Entropy of the Instantons}
\label{action}
We now turn to the problem of calculating the action of these
instantons. These calculations will mostly be done within the context of
the quasi-local formulation for gravitational thermodynamics that has been
developed by Brown and York \cite{BY}. For simplicity we will
only consider the lukewarm and cold Kerr deSitter instantons. A complete 
discussion for all three types of instantons
including electromagnetic fields, corner terms, and reference spacetimes  
will be presented in \cite{tbp}. Here however, we shall just present the
general ideas.

Let us begin by considering a region of spacetime $M$ bounded by two
time-like boundaries $B_i$ and $B_o$, and two spacelike boundaries
$\Sigma_i$ and $\Sigma_f$. Further we foliate the spacetime regions with
spacelike hypersurfaces such that the initial and final hypersurfaces
coincide with $\Sigma_i$ and $\Sigma_f$ respectively. Specializing to our
situation, the timelike boundaries will be the ($r= \mbox{constant}$)
hypersurfaces just outside/inside the black hole/cosmological horizons,
while the spacelike foliations/boundaries will be ($t =
\mbox{constant}$) hypersurfaces. With this region in mind we turn to
considering the action.

As mentioned in the introduction, before a calculation of the actions can
be made, it is necessary to decide what specific action is to be
calculated. A variety
of actions may be chosen that, upon extremization, will yield the
Einstein field equations in the interior of a spacetime region,
M. They will differ however, in what quantities will be required to
remain fixed on the boundaries $\partial M$ of said region. 
As a concrete and pertinent example consider the usual
action.
\bea
\label{a1}
I_{Stan} &=& \frac{1}{16\pi} \int_M d^4 x \sqrt{-g} \left( \cR - 2 \Lambda
\right)
 \\ &+& \frac{1}{8\pi} \int_{\Sigma_f - \Sigma_i} d^3x \sqrt{h} K -
\frac{1}{8\pi} \int_{B_o - B_i} d^3x \sqrt{-\gamma} \Theta,
\eea  
where $g$ is the determinant of the spacetime metric $g_{ab}$ and $h$ and
$\gamma$ are the determinants of the induced boundary metrics. $\cR$ is 
the Ricci scalar, $\Lambda$ is the cosmological constant, and $K$ and
$\Theta$ are the traces of the extrinsic curvatures of the boundaries. The
variation of this action is given by,
\bea
\delta I_{Stan} &=& \frac{1}{16\pi} \int_M d^4 x \sqrt{g} \left( G_{ab} +
\Lambda g_{ab} \right) \delta g^{ab} + \int_{\Sigma_f - \Sigma_i} d^3x
P^{ab} \delta h_{ab} \\ &+& \int_{B_o - B_i} d^3x
\Pi^{ab} \delta \gamma_{ab} ,
\eea
where $P^{ab} = \frac{1}{8 \pi} \sqrt{h} \left( K h^{ab} -
K^{ab} \right)$, and $\Pi^{ab} = \frac{1}{8 \pi} \sqrt{-\gamma} \left(
\Theta h^{ab} - \Theta^{ab} \right)$.
In this case, it can be seen that the choice of action requires
that the components of the induced boundary metric be fixed on the
boundaries.

To investigate the relevance of this action to our situation, let us now
perform the standard ADM \cite{ADM} analysis, as
extended by Brown and York \cite{BY}, to get the following form for the
variation of the action,
\bea
\delta I_{Stan} &=& \left( \mbox{terms that are zero for solutions of the
field equations} \right) \\ 
&+& \int_{\Sigma_f-\Sigma_i} d^3 x P^{ab}\delta
h_{ab} + \int_{B_{o}-B_{i}} d^3 x \sqrt{\sigma} \left( -\varepsilon_E
\delta N + j_a \delta V^a + \frac{N}{2} s^{ab} \delta \sigma_{ab} \right).
\eea 
In the above, $\sigma_{ab}$ is the induced metric on the intersection of
timelike boundary with the spacelike foliations, 
$\varepsilon_E$, $j_a$, and $s_{ab}$ are respectively a
quasilocal energy
density, a quasilocal momentum density, and a quasilocal stress density 
defined on those boundaries, while $N$ and $V^a$ are respectively the
lapse and shift functions, also defined on those boundaries.

Now, only some of these boundary conditions are correct for the work that
we
are doing. Ultimately we are going calculate the action 
of a particular instanton, and the
boundary terms will define what ensemble we are studying (recall
from the introduction, that the instanton gives a first
approximation to the full path integral calculation involving all possible
interpolating
spacetimes). We will be interested in the microcanonical ensemble, 
and so will want boundary conditions that fix the quasilocal
energy  and momentum densities. On the other hand however, we will 
want the induced spatial metrics on the spatial boundaries to be
fixed so that the instantons may be cleanly matched to the
Lorentzian solutions. Finally, removal of conical singularities requires a
fixing of the surface
stress tensor density, $s_{ab}$. With these considerations in mind, 
it is a trivial matter to modify the action so that the conditions will be
fulfilled. The new action is,
\be
\label{caction}
I_{Micro} = I_{Stan} + \int_{B_o - B_i} d^3 x \sqrt{\sigma} \left( N
\varepsilon_E - V^a j_a - \frac{N}{2} s^{ab}\sigma_{ab} \right).
\ee

We may now calculate instanton actions - or rather the actions of
$\frac{1}{2}$ of the instantons since it is only half instantons that we
connect to the Lorentzian spacetime (figure \ref{fig2}). Performing the
calculations, we obtain, from (\ref{caction})
\bea
I_{Micro-Lukewarm} &=& \frac{1}{8} \left( \cA_{BH} + \cA_C \right) \\
I_{Micro-Cold} &=& \frac{1}{8} \cA_{C},
\eea
where $\cA_{BH}$ is the area of the black hole horizon, and $\cA_C$ is the
area of the black hole horizon.

Now, having chosen microcanonical boundary conditions the partition
function $Z = \Psi^2$ can be interpreted as the density of states, and
therefore the entropy of the spacetimes is $\ln Z$ \cite{rob}. Thus, this
calculation
gives the standard results for entropies of spacetimes. Entropy is one
quarter of the sum of the areas of the horizons for the luke warm case,
and one quarter of the area of the cosmological horizon for the cold case.

The pair creation rates are given by,
\be
\Gamma = \Psi^2 \approx e^{-2I}.
\ee
Thus we see that the rates are always suppressed by an amount equal to the
entropy of the created spacetime.

\section*{Acknowledgements}
The authors would like to thank the Natural Sciences and Engineering
Council of Canada for financial support. We gratefully acknowledge correspondence
by R. Bousso and S. Ross.

\end{document}